# Planar He plasma jet: plasma "bullets" formation, 2D "bullets" concept and imaging

Danil Dobrynin and Alexander Fridman

*C&J Nyheim Plasma Institute, Drexel University, Camden NJ 08103*

*Abstract*

Applications of dielectric barrier discharge (DBD) based atmospheric pressure plasma jets are often limited by the relatively small area of treatment due to their 1D configuration. This letter describes the first results demonstrating generation of 2D plasma "bullets" and 2D plasma jets permitting fast treatment of large targets. Imaging of pulsed DBD in He and ionization wave propagation along the dielectric surface show that DBD evolution starts with formation of transient anode glow, and only then continues with development of cathode-directed streamers. The anode glow can propagate as an ionization wave along the dielectric surface outside of the discharge gap. We show that plasma "bullets" propagation is not limited to 1D geometry (tubes), and can be organized in a form of a planar plasma jet, or other 2D (or even 3D) shapes.

Non-thermal atmospheric plasma jets (APPJ) based on dielectric barrier discharge (DBD) have attracted considerable interest in the recent decades. [1-3]. Multiple research groups have studied these plasma sources for various applications like surface modification, [5, 6] decontamination, [6-9] cancer treatment [10-12] and others, as well as their thorough diagnostics to understand chemical and physical characteristics. One of the important challenges and limitations of existing 1D plasma jets is related to relatively low area of treatment.[13-19] The main focus of this letter is to address this problem and to offer possibility to generate 2D plasma jets that allow fast treatment of large surface area targets.

Many reports have shown that plasma jets are in fact propagating via so-called "plasma bullets", or a train of discrete surface ionization waves traveling along dielectric surface (sometimes to address the ionization wave reconnection to the "parent" DBD plasma, they are called "pulsed atmospheric-pressure plasma streams (PAPS)"). Although the physics of 1D jets and plasma "bullets" propagation has been described, [20-30] connection between the processes in the DBD plasma and "bullet" initiation is still not clearly understood. Here we demonstrate that plasma "bullets" originate from the initial anode glow (surface discharge) in DBD that precedes volumetric streamer discharge and propagates along the surface of dielectric to later form the jet upon exiting outside of the discharge chamber into the atmosphere. Based on this observation, we show possibility of rearranging the classical 1D tube configuration of APPJ into a 2D plasma jet.

In these studies, we have used microsecond-pulsed high voltage power supply with output of 20 kV peak-to-peak, pulse duration of ~8 µs (Figure 1, a), and frequency of 500 Hz.[31] Plasma jets were generated in He flow (99%, Airgas). To monitor the discharge development and ionization waves propagation we have used 4Picos ICCD camera from Stanford Computer Optics triggered using P6015A high-voltage probe (75-MHz bandwidth, Tektronix) also connected to a 1-GHz DPO-4104B oscilloscope (Tektronix).

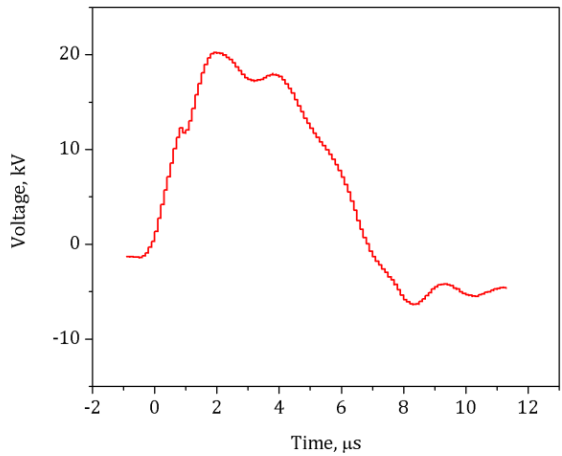

*(a)*

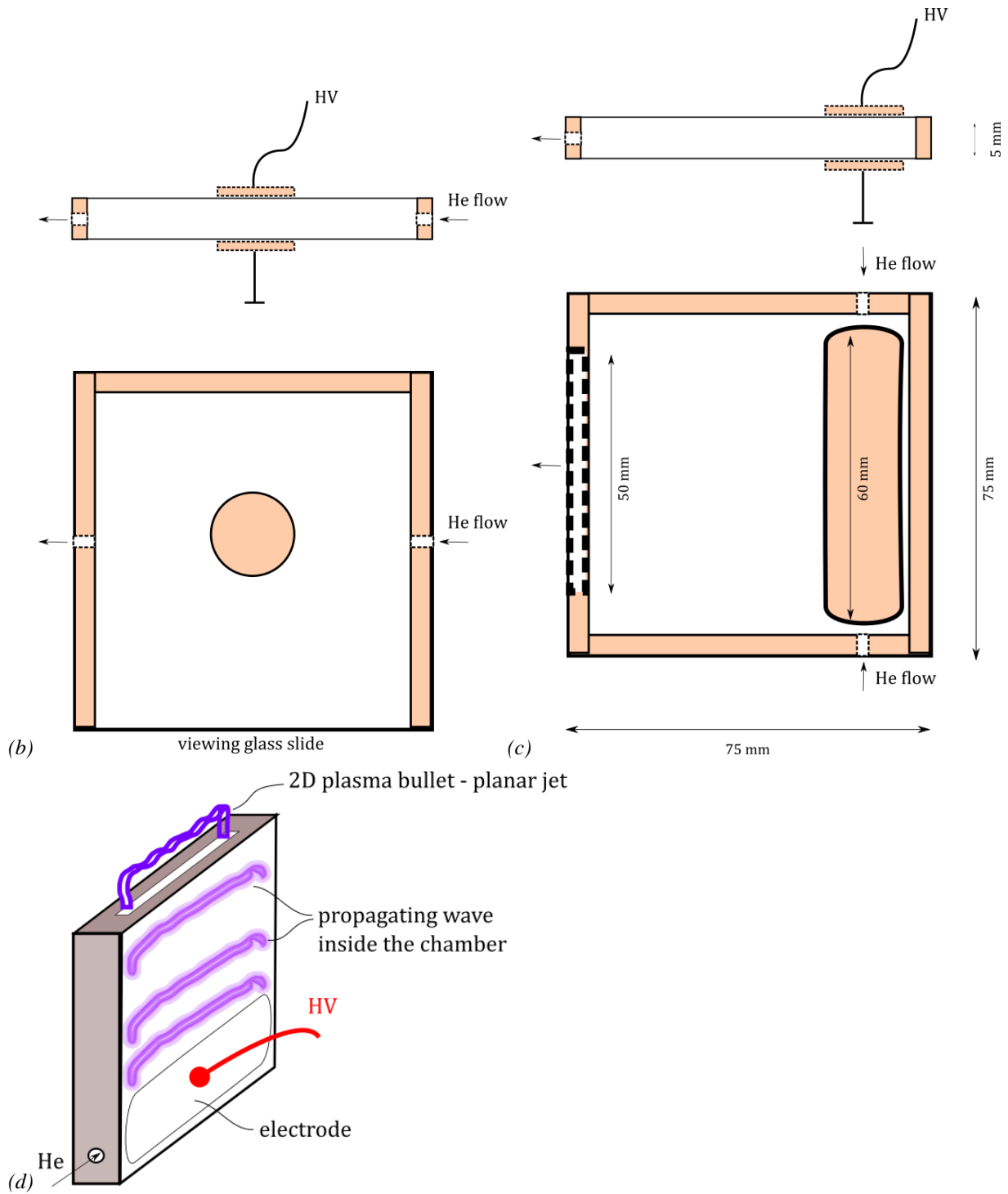

*Figure 1.* Experimental setup: (a) Typical voltage waveform; (b) Discharge chamber schematic for visualization of surface ionization wave development and propagation; (c) Discharge chamber schematic for generation of planar plasma jet; and (d) Drawing of the planar plasma jet setup

Two similar discharge chambers were used for generation and imaging of DBD, surface ionization waves, plasma "bullets" and planar plasma jet. Both chambers were made of two 1 mm thick 75x75 mm glass slides separated by 5 mm thick dielectric to form a rectangular cuboid. The first setup used to study DBD development and propagation of surface ionization waves employed a 1 mm thick glass slide glued on the side of the chamber to visualize the discharge from the side (Figure 1, b). Here we used two 20 mm in diameter electrodes fixed in the center of glass sides of the chamber, as well as two 2 mm holes for He flow though the setup at ~0.5 L/m. To demonstrate possibility of reconfiguring the 1D APPJ into a planar plasma jet, in the second chamber instead of a circular gas outlet we have used a 1.5 mm wide and 50 mm long slit, as well as 60 mm long and 20 mm wide copper electrodes fixed on the outer surfaces of the glass slides (Figure 1 c, d). Gas was supplied through two 2 mm holes at rate of ~5 L/m.

Our first experiments aimed to understand the development of plasma bullets. For that, we have employed fast imaging technique using ICCD camera triggered from the high voltage pulse. Since during the high voltage pulse the discharge reignites several times (a series of breakdowns), we have focused only on the first appeared illumination. At first, the DBD starts with development of a series of avalanches traveling from the grounded electrode towards powered anode. As the avalanches reach the anode, presence of the dielectric causes accumulation of the charge lading to generation of electron densities (and local electric fields) sufficient for formation of surface discharge that appears as a bright anode glow region (Figure 2). This anode glow – "pancake" – is generated prior to the volumetric discharge that develops via traditional cathode-directed streamers. This series of events is not limited to the case of He atmosphere, in fact, similar discharge behaviors can be seen also for atmospheric air nanosecond-pulsed DBD (see, for example, [32]).

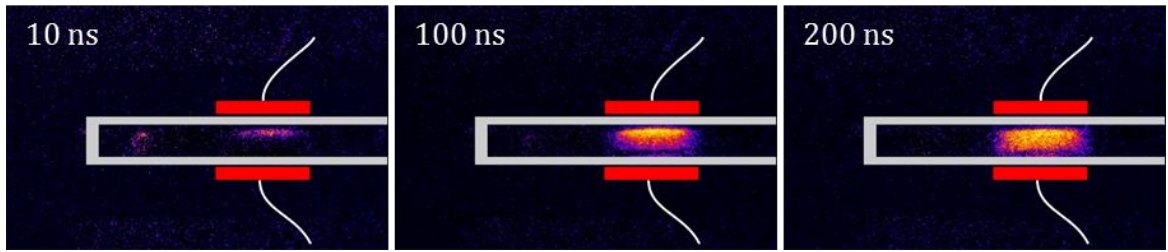

*Figure 2*. Development of DBD in He – single exposure ICCD images (false color) taken with 10 ns exposure time and indicated delay time after the trigger (when high voltage pulse reaches ~5 kV). Faint glow seen on the first image in the left region of the chamber is an imaging artefact.

Because the anode glow region can be viewed as a second capacitor (third "electrode" near the anode), high electric fields at its edges facilitate development of surface wave propagating along the surface of the dielectric. This phenomenon is clearly seen on the Figure 3 as a radially propagating ring with the velocity of ~30 km/s – these images were taken from both side and top of the discharge chamber. As the wave reaches the output gas hole in the dielectric wall, it exits the discharge chamber in a form of a plasma "bullet". In this experiment, because of the discharge chamber configuration the "bullet" is most probably does not have a traditional "donut" shape – unlike in tubular setups, where the ionization wave propagates in the tube along its inner surface. [3, 33, 34]

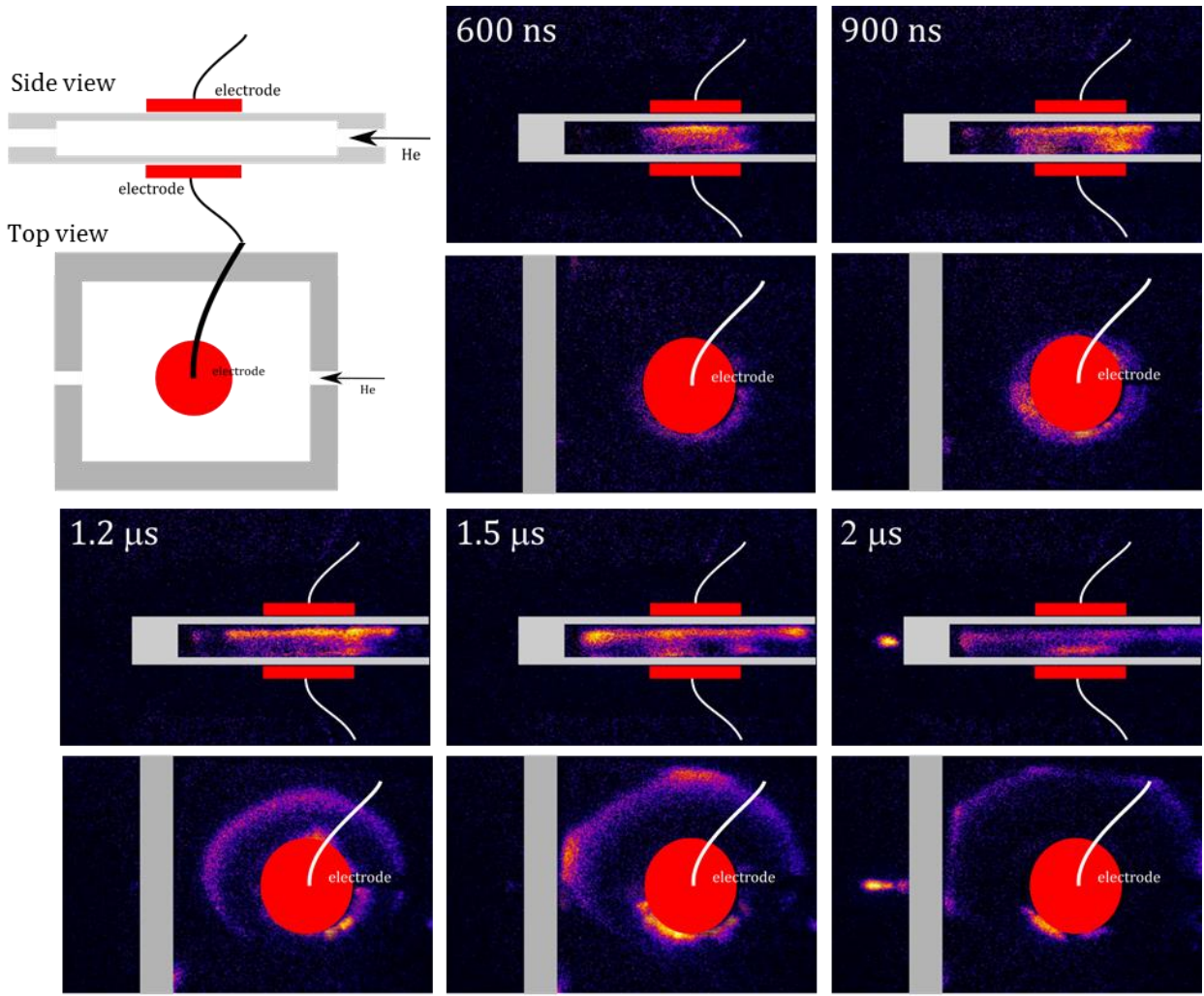

*Figure 3.* Propagation of surface ionization wave – single exposure ICCD images (false color) taken with 300 ns exposure time and indicated delay time after the trigger (when high voltage pulse reaches ~5 kV). Images are taken separately for the top and side views.

Here we have demonstrated the relation between formation of surface ionization wave in DBD and plasma "bullets". Because in this configuration we are able to generate a 2D ionization surface wave, we propose a new configuration of APPJ to generate planar plasma jets. For that we have used the second discharge chamber with exit gas opening in a form of a long slit. In these experiments, similar to the first case, one can clearly see propagation of ionization wave inside the chamber along the dielectric surface (Figure 4). As the wave reaches the exit opening slit it forms planar plasma jet propagating into the outer atmosphere, however due to flow conditions that cause to effective mixing with air, and possibly low specific energy input, propagation length of the jet is only ~1 cm or less. Because, in principle, here the output gas opening is not limited to a straight planar slit and can be organized in a more complex shape (for example, wave-shaped slit), it is possible to generate plasma jets of various complex 3D profiles. Similarly, the width of the jet can be controlled by the distance from the electrodes and, in general, limited only by the pulse energy provided by the power supply. It is possible, therefore, to generate jets with any width to accommodate treatment of large targets.

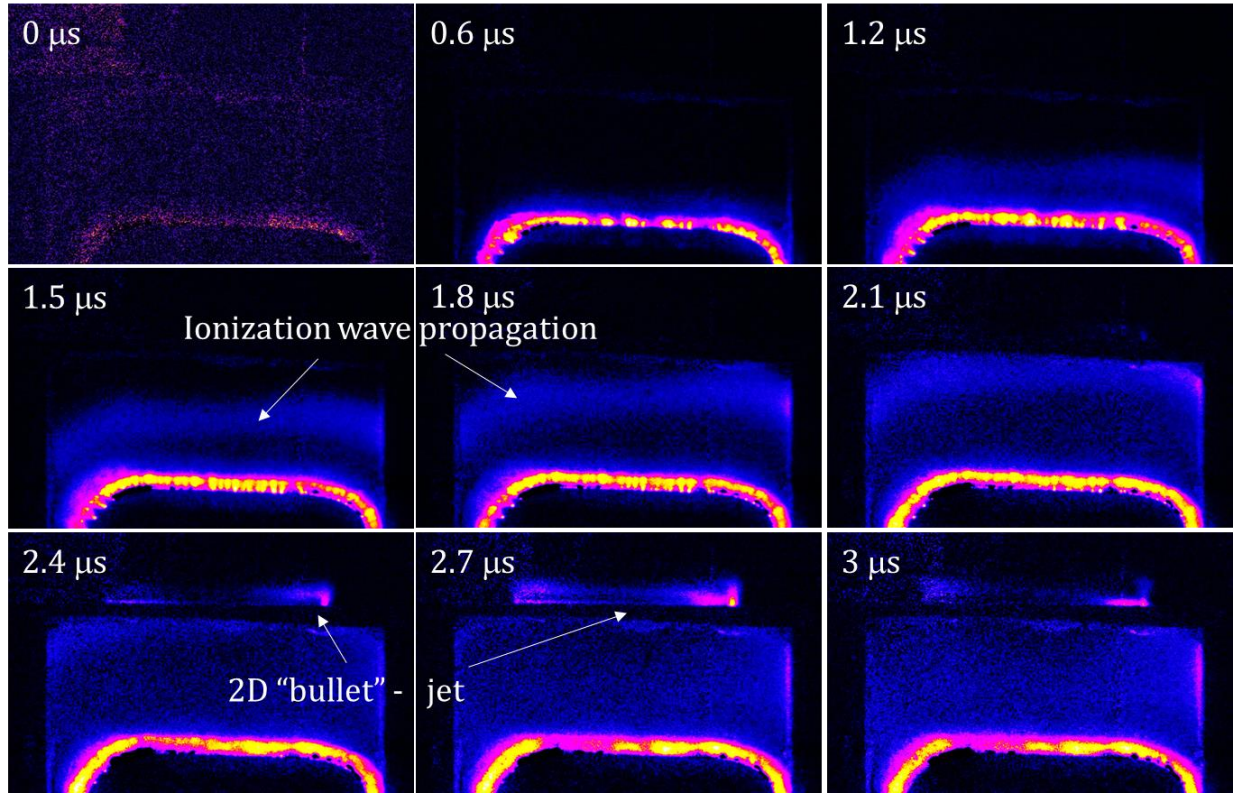

*Figure 4.* Propagation of surface ionization wave and formation of planar plasma jet – single exposure ICCD images (false color) taken with 300 ns exposure time and indicated delay time after the trigger (when high voltage pulse reaches ~5 kV).

Photographs shown in Figure 5 demonstrate possible applications for treatment – attachment of the jet to a floating insulated wire, and a finger. One may notice that although the distance from the discharge chamber output opening to the target is not constant and varies from ~0.5 – 1 cm, the plasma jet allows relatively uniform treatment. Also, because of the conductance of the tissue (floating potential), plasma jet can propagate to longer distance. More detailed studies are clearly needed to explore this phenomena and 3D plasma jet generation method in details.

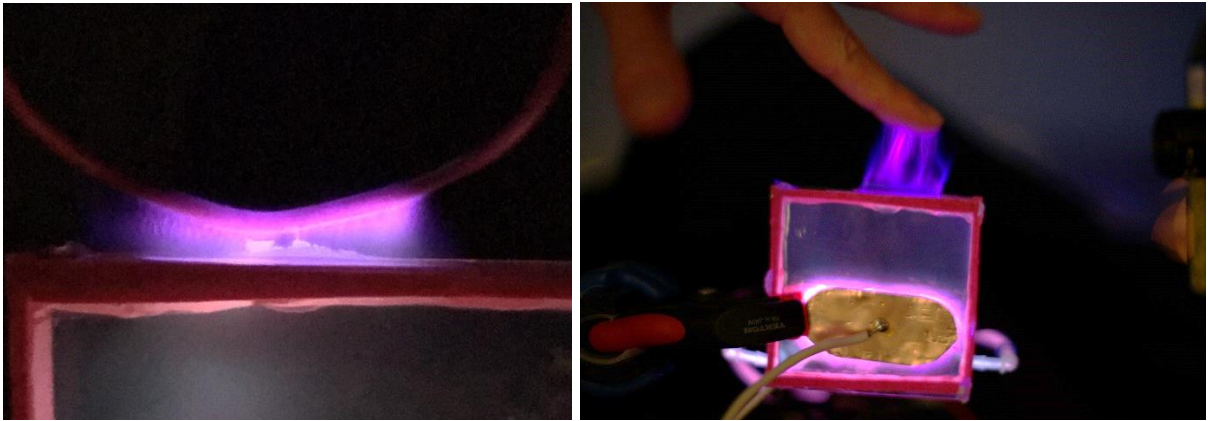

*Figure 5*. Photographs of planar plasma jet applications to a floating potential insulated wire and a finger.

In conclusion, imaging of pulsed DBD in He and ionization wave propagation along the dielectric surface brings us to the following conclusions:

1) Evolution of pulsed DBD plasma starts with formation of transient anode glow, and only then continues with development of cathode-directed streamers;
2) In the case of noble gases, the anode glow can propagate as ionization wave along the dielectric surface outside of the discharge gap. For the case of traditional dielectric tubes, the ionization wave appears as what is known as a "plasma bullets", often with characteristic "donut" shape;
3) Plasma "bullets" propagation is not limited to 1D geometry (tubes), and can be organized in a form of a planar jet, or other 2D (or even 3D) shapes.
4) Propagation of the 2D "bullets" results in generation of 2D (or even 3D) plasma jets. The size of these jets is limited by the pulse energy and gas flow characteristics.

Plasma jets with complex 2D (or 3D) geometries open new possibilities in the areas of large surface material processing and plasma medicine

**Acknowledgment**

This work was partially supported by the NSF/DOE Partnership in Basic Plasma Science and Engineering grant (DOE grant DE-SC0016492, PI: D. Dobrynin).